\documentclass[10pt]{iopart}
\usepackage[dvips]{color}
\usepackage{Graphen}
\usepackage{graphicx}
\begin{document}
\title[Orbital order in degenerate Hubbard models]{Orbital order in
  degenerate Hubbard models : A variational study}

\author{J.~B\"unemann${}^1$, K.~J\'avorne-Radn\'oczi${}^2$, 
P.~Fazekas${}^1$\footnote{During the final stage of proofreading for 
this work Patrik Fazekas deceased on May 16, 2007.}, 
and F.~Gebhard${}^1$ }

\address{${}^1$Fachbereich Physik, Philipps-Universit\"at Marburg,
D--35032 Marburg, Germany \\
${}^2$Research Institute for Solid-State Physics and Optics,
Hungarian Academy of Sciences, H-1525 Budapest, Hungary}
\begin{abstract}
We use the Gutzwiller variational many-body theory to investigate the 
stability of orbitally ordered states in a two-band Hubbard-model without 
spin degrees of freedom. Our results differ significantly from earlier 
Hartree-Fock calculations for this model. 
The Hartree-Fock phase diagram displays a large variety of orbital orders. 
In contrast, in the Gutzwiller approach orbital order only appears for 
densities in a narrow region around half filling.      
\end{abstract}



\section{Introduction}
\label{intro}
The investigation of orbital degrees of freedom has become an important
field in theoretical solid-state physics over the past two decades. There are
a number of materials which are believed to show phase transitions with 
orbital order parameters. Among them, the perovskite manganites, 
e.g., ${\rm La}_{1-x}{\rm Sr}_x {\rm MnO}_3$, have attracted particular 
interest because of the colossal magnetoresistance behaviour  which is 
observed in these materials. In theoretical 
studies on manganites, one often neglects the almost localised 
Mn $t_{2 {\rm g}}$ orbitals and investigates solely the electronic properties 
of a systems with two $e_{\rm g}$ orbitals per lattice site. 
 In order to study the ferromagnetic 
 phase of such a model, Takahashi and Shiba \cite{Takahashi}  further 
neglected the spin-degrees of freedom because the Hund's rule coupling 
is assumed to align the spins in the two $e_{\rm g}$~orbitals. 
 
 The mean-field study in  \cite{Takahashi} found a surprisingly large
number of stable orbitally ordered phases for the spinless $e_{\rm g}$ model. 
However, it is well known that mean-field approximations tend to 
overestimate the stability of ordered phases in correlated electron systems. 
Therefore, the purpose of this work is to reinvestigate the two-orbital 
 Hubbard model  without spin-degrees of freedom by means of the
 Gutzwiller variational many-body theory.    

Our paper is organised as follows: The Hamiltonian 
and the different types of order parameters that we are going to investigate 
are discussed in section~2. In section~3 we introduce Gutzwiller 
wave functions and derive an approximate expression for the variational 
ground-state energy. Numerical results are presented in section~4 and a 
brief summary closes our presentation in section~5.  

\section{Model system and types of order parameters}
\label{sec:Hamilt}
We investigate a two-orbital ($e_{\rm g}$-type) Hubbard model \cite{Hubbard}
 without
spin-degrees of freedom, 
\begin{eqnarray}
  \label{1}
\hat{H}&=&\hat{H}_{0}+\hat{H}_{U} \,,\\ \label{1b}
\hat{H}_{0}&\equiv&\sum_{i,j}\sum^2_{b,b'=1}t^{b,b'}_{i,j}\hat{c}^{\dagger}_{i,b}
\hat{c}^{}_{j,b'}\,, \\\label{1c}
\hat{H}_{U}&\equiv&U\sum_{i}\hat{n}_{i,1}\hat{n}_{i,2}\, .
\end{eqnarray}
Here, the tight-binding parameters $t^{b,b'}_{i,j}$
describe hopping processes between orbitals $b,b'$ on cubic
lattice sites ${\bf R}_{i}$ and ${\bf R}_{j}$, respectively. The Hamiltonian
(\ref{1}) is formally equivalent to the standard one-band Hubbard model if
the indices $b,b'$ are regarded as spins. However, the tight-binding
parameters
in (\ref{1b})
would be unusual for a genuine one band model since they contain
inter-orbital hopping-terms. The Hubbard parameter $U$
in (\ref{1c}) is derived from  $U=U'-J$ where $U'$ and $J$ are the Coulomb
and the exchange interaction between electrons in different $e_{\rm g}$
orbitals. 

\begin{figure}[htb]
\centerline{\includegraphics[clip,width=7cm]{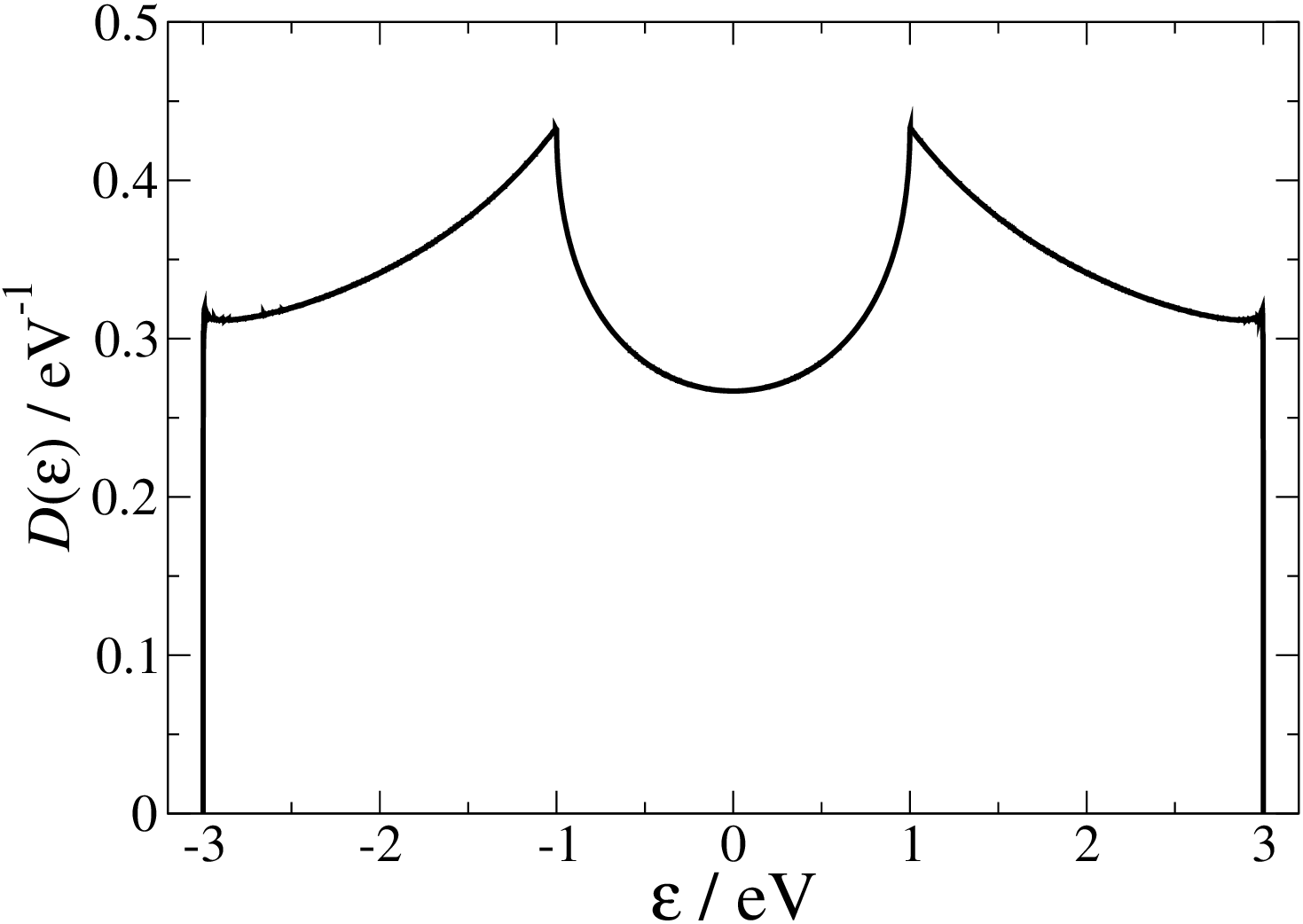}}
\caption{Density of states of the model system (\ref{2}) .}\label{fig1}
\end{figure}

We restrict our investigation to systems with only nearest neighbour
hopping since additional hopping terms would only 
destabilise the orbital order we are interested in. For $e_{\rm g}$ orbitals 
$|z_1\rangle\equiv|x^2-y^2 \rangle$,
$|z_2\rangle\equiv|3z^2-r^2\rangle$ and
hopping parameters $t_{dd\sigma}=1 {\rm eV}$, 
$t_{dd\delta}=0 {\rm eV}$ \cite{SlaterKoster}, 
the one-particle Hamiltonian in momentum space reads 
\begin{equation}  \label{2}
\hat{H}_{0}=
\sum_{{\bf k}}\sum_{b,b'}\epsilon^{b,b'}_{z;{\bf k}}
\hat{c}^{\dagger}_{z;{\bf k},b}
\hat{c}^{}_{z;{\bf k},b'}\,,
\end{equation}
with \cite{SlaterKoster}
\begin{eqnarray} \nonumber
\epsilon^{1,1}_{z;{\bf k}}&=&(\cos{(k_x)}+\cos{(k_y)}+
4\cos{(k_z)})/2\, ,\\
\epsilon^{2,2}_{z;{\bf k}}&=&3(\cos{(k_x)}+\cos{(k_y)})/2\, ,\\ \nonumber
\epsilon^{1,2}_{z;{\bf k}}&=&-\sqrt{3}(\cos{(k_x)}-\cos{(k_y)})/2=
\epsilon^{2,1}_{z;{\bf k}}\, .
\end{eqnarray}
 The ${\bf k}$-integrated density of states $D(\varepsilon)$ that results from 
this band structure is shown in figure~\ref{fig1}. 

For our investigation of orbital order we need to introduce two
more basis representations of the orbital space,
\begin{eqnarray} \label{4}
  |x_1\rangle&\equiv&\frac{1}{\sqrt{2}}(|z_1\rangle +
  |z_2\rangle)\,,\\  \nonumber
  |x_2\rangle&\equiv&\frac{1}{\sqrt{2}}(|z_1\rangle - |z_2\rangle)\,,
\end{eqnarray}
and 
\begin{eqnarray}\label{5}
  |y_1\rangle&\equiv&\frac{1}{\sqrt{2}}(|z_1\rangle + \rm{i}
  |z_2\rangle)\,,\\  \nonumber
  |y_2\rangle&\equiv&\frac{1}{\sqrt{2}}(|z_1\rangle - \rm{i}|z_2\rangle)\,.
\end{eqnarray}
 The dispersion relations in momentum space for the `$x$'- and the 
`$y$'-basis  become
\begin{eqnarray} 
\epsilon^{1,1}_{x;{\bf k}}&=&((2-\sqrt{3})\cos{(k_x)}+(2+\sqrt{3})\cos{(k_y)}+
2\cos{(k_z)})/2\, ,\\ \nonumber
\epsilon^{2,2}_{x;{\bf k}}&=&((2+\sqrt{3})\cos{(k_x)}+(2-\sqrt{3})\cos{(k_y)}+
2\cos{(k_z)})/2 ,\\ \nonumber
\epsilon^{1,2}_{x;{\bf k}}&=&\cos{(k_x)}+\cos{(k_y)}-2\cos{(k_z)})/2
=\epsilon^{2,1}_{x;{\bf k}}\, ,\\ 
\epsilon^{1,1}_{y;{\bf k}}&=&\cos{(k_x)}+\cos{(k_y)}+\cos{(k_z)}
=\epsilon^{2,2}_{y;{\bf k}}  \, ,\\ \nonumber
\epsilon^{1,2}_{y;{\bf k}}&=&
((1+\sqrt{3}\rm{i})\cos{(k_x)}+(1-\sqrt{3}\rm{i})
\cos{(k_y)}-2\cos{(k_z)})/2 \, ,\\ \nonumber
\epsilon^{2,1}_{y;{\bf k}}&=&\left(\epsilon^{1,2}_{y;{\bf k}}\right)^{*}\, .
\end{eqnarray}
The interaction term (\ref{1c}) has the same form 
for all three basis representations $\xi=x,y,z$.
  
We describe orbital order in our model system (\ref{1}) through the
  parameters
\begin{eqnarray}
\tau_{\xi;i}&\equiv&(\langle \hat{n}_{i,\xi_1}  \rangle -
\langle \hat{n}_{i,\xi_2}  \rangle)/2 \,,\\
\hat{n}_{i,\xi_b}&\equiv&\hat{c}^{\dagger}_{i,\xi_b}\hat{c}^{}_{i,\xi_b} \,,
\end{eqnarray}
for each of the three representations $\xi$. By using 
the  Pauli matrices $\tilde{1},\tilde{\tau}^{x},
\tilde{\tau}^{y},\tilde{\tau}^{z}$ the order parameter can also be  
written  as (compare reference \cite{Takahashi})
\begin{equation}
\tau_{\xi;i}=\frac{1}{2}\sum_{b,b'}\langle \hat{c}^{\dagger}_{i,z_b}
 (\tilde{\tau}^{\xi})_{b,b'}\hat{c}^{}_{i,z_{b'}}\rangle\,.
\end{equation}

Besides the orbital character of the order-parameter we need to specify its
lattice site dependence. Following reference \cite{Takahashi} we consider 
orders of the form 
\begin{equation}
 \label{10}
\tau_{\xi;i}=\tau_{\xi,0}\exp{({\rm i}{\bf Q}{\bf R}_{i})}
\end{equation} 
with commensurate vectors ${\bf Q}$ which belong to the $\Gamma$ point 
(${\bf Q}=(0,0,0)$), the $R$ point (${\bf Q}=(\pi,\pi,\pi)$), 
the $X$ point (${\bf Q}=(0,0,\pi)$), and the $M$ point
(${\bf Q}=(\pi,\pi,0)$). The real parameter $\tau_{\xi,0}$ in (\ref{10}) is
independent of the lattice site vector ${\bf R}_{i}$ and assumed to be 
positive. Note, that for vectors ${\bf Q}\neq(0,0,0)$ 
equation (\ref{10}) divides the lattice into an `$A$'-lattice with a 
majority $\xi_1$ occupation ($\tau_{\xi;i}>0$), and a `$B$'-lattice 
with a majority $\xi_2$ occupation ($\tau_{\xi;i}<0$).

\section{Gutzwiller wave-functions}
\subsection{Definitions}
For an investigation of the Hamiltonian (\ref{1}) we use Gutzwiller 
variational wave functions \cite{Gutzwiller}  which are defined as 
\begin{equation}\label{15}
| \Psi_{\rm G}\rangle \equiv \prod_{i} 
 \hat{P}_{i} | \Psi_0\rangle \; .
\end{equation}
Here, $| \Psi_0\rangle$ is a normalised quasi-particle vacuum and the 
local correlator has the form
\begin{equation}
\label{16}
\hat{P}_{i}=\sum_{I}\lambda_{i,I} \hat{m}_{i,I}
\end{equation}
where $\hat{m}_{I}=| I\rangle \langle I|$ projects onto the four 
local configuration states 
$| I\rangle$, i.e., the empty state $|\emptyset\rangle$, 
the doubly occupied state $|d\rangle$ and the two single electron states 
$|\xi_1\rangle,|\xi_2\rangle$. Note,
 that  $|\xi_1\rangle$ and $|\xi_2\rangle$
depend on the orbital representation $\xi$ whereas the states 
$|\emptyset\rangle$ and $|d\rangle$ are invariant
under the orbital transformations (\ref{4}) and (\ref{5}). For the 
variational parameters  $\lambda_{i,I}$ we make an Ansatz which is 
consistent with the spatial symmetry of the order parameter,
\begin{eqnarray}
\lambda_{i,\emptyset}&=&\lambda_{\emptyset}\,,\\ 
\lambda_{i,d}&=&\lambda_{d}\,,\\
\lambda_{i,\xi_1}&=&\lambda_{s}+\delta\lambda_{s}
\exp{({\rm i}{\bf Q}{\bf R}_{i})}\,,\\
\lambda_{i,\xi_2}&=&\lambda_{s}-\delta\lambda_{s}
\exp{({\rm i}{\bf Q}{\bf R}_{i})}\,,
\end{eqnarray}
where the parameters $\lambda_{\emptyset}$,$\lambda_{d}$,
$\lambda_{s}$,$\delta\lambda_{s}$ are independent of the lattice site 
vector ${\bf R}_{i}$.

A more general ansatz for the Gutzwiller correlator $\hat{P}$ 
which also includes non-diagonal configuration transfer operators
$| I\rangle \langle I'|$ has been studied in reference \cite{springer}. 
As we will see below (section \ref{3.2}), it is sufficient 
for our calculations in this work to consider 
only diagonal operators in $\hat{P}_{i}$.   

\subsection{Evaluation in infinite dimensions}\label{3.2}
For any practical use of a variational wave-function it is 
essential that the expectation value of the Hamilton  can be 
calculated. However, despite the simplicity of the Gutzwiller wave-function
the evaluation of
\begin{equation}
E_{\rm var}=\langle \hat{H}  \rangle_{\Psi_{\rm G}}
\end{equation}
poses a difficult many-particle problem that 
cannot be solved in general. Gutzwiller introduced an approximate 
evaluation scheme that was based on quasi-classical counting
arguments. More recent derivations of this approximation
can be found in references \cite{RMPVollhardt,JBcounting}. Analytically
exact evaluations were later found to be possible in one dimension
\cite{Gutzwiller1dim,Gutzwiller1dimb,Kollar}, and in infinite spatial dimension
\cite{florian}. The results of the latter evaluation 
turned out to be equivalent to the Gutzwiller approximation
for systems which can be studied within this approach. The 
evaluation scheme in infinite dimensions was later generalised for the 
investigation of multi-band Hubbard models \cite{PRB}  
and superconducting systems \cite{springer,supra}. 
We will use these exact results in infinite dimensions as an approximation 
in order to evaluate expectation values of our Hamiltonian (\ref{1}). 

We only consider single-particle wave functions $|\Psi_0\rangle$ in
 (\ref{15})
for which the local density matrix $\tilde{C}^{0}$ with the matrix 
elements
\begin{equation}
  C^{0}_{i,\xi_b,\xi_{b'}}\equiv\langle
  \hat{c}^{\dagger}_{i,\xi_b}\hat{c}^{}_{i,\xi_{b'}} \rangle_{\Psi_0}
\equiv n^{0}_{i,\xi_b}
\delta_{b,b'}
\end{equation}
is diagonal with respect to $b,b'$. Finite 
non-diagonal elements in $\tilde{C}^{0}$ could only appear if we were
mixing different order parameters. 

As shown in \cite{springer} the four parameters $\lambda_{i,I}$ for a
lattice site $i$ have to obey the constraints
 \begin{eqnarray}
\label{30z}
1&=&\langle\hat{P}_{i}^2 \rangle_{\Psi_0} \;, \\ \label{30b}
C^{0}_{i,\xi_b,\xi_{b'}}&=&\langle\hat{P}^2_{i} 
 \hat{c}^{\dagger}_{i,\xi_b}\hat{c}^{}_{i,\xi_{b'}}
  \rangle_{\Psi_0}\;.
\end{eqnarray}
Our correlation operator (\ref{16}) automatically fulfils 
the constraints (\ref{30b}) for $b\neq b'$. 
This is the reason why it was allowed in the first place
 to include only diagonal operators $\hat{m}_{i,I}$  in  (\ref{16}). 
Consequently, instead of (\ref{30b})
 we only need to consider the diagonal constraints
\begin{equation}
n^{0}_{i,\xi_b}=\langle \hat{P}^2_{i} \hat{n}_{i,\xi_b}
\rangle_{\Psi_0}\;.
\end{equation}
All constraints can be solved explicitly if we use the 
results for local expectation values 
\begin{equation}
\label{34}
m_{i,I}\equiv \langle  \hat{m}_{i,I}  \rangle_{\Psi_{\rm G}}
=\lambda_{i,I}^{2}\langle  \hat{m}_{i,I}  \rangle_{\Psi_{0}}
\end{equation}
which hold in infinite spatial dimensions. With (\ref{34})
we can use the expectation values $m_{i,I}$ as new variational 
parameters instead of $\lambda_{i,I}$.
 The constraints then read
\begin{eqnarray}
1&=&m_{i,\emptyset}+m_{i,d}+m_{i,\xi_1}+m_{i,\xi_2}\,,  \\
n^{0}_{i,\xi_b}&=&m_{i,\xi_b}+m_{i,d}\,.
\end{eqnarray}
and can be readily solved by expressing all local occupancies in terms
of the average numbers of doubly occupied sites $m_{i,d}$,
\begin{eqnarray}
m_{i,\emptyset}&=&1-n^{0}_{i,\xi_1}-n^{0}_{i,\xi_2}+m_{i,d}\,,  \\ \label{27}
m_{i,\xi_b}&=&n^{0}_{i,\xi_b}-m_{i,d}\,.
\end{eqnarray}
Apart from the still unspecified
one particle wave function $|\Psi_0 \rangle$ and the corresponding 
local density matrices $\tilde{C}^0_i$ the probabilities  $m_{i,d}$ are
the only remaining variational parameters. Note, that (\ref{27}) leads to 
\begin{equation}
n_{i,\xi_b}
\equiv\langle  \hat{n}_{i,\xi_b} \rangle_{\Psi_{\rm
    G}}=n^{0}_{i,\xi_b}
\end{equation}
for the orbital densities in the correlated Gutzwiller wave-function. 
We skip the explicit declaration of the  orbital representation $\xi$ 
for the rest of this chapter and write $b$ instead of $\xi_b$ in all indices. 

In infinite dimensions the expectation value
of the one particle Hamiltonian (\ref{1b}) becomes \cite{PRB}
\begin{eqnarray}
\label{28}
 E_0=\langle \hat{H}_0  \rangle_{\Psi_{\rm G}}&=&
\sum_{i,j}\sum_{b,b'}\sqrt{q_{i,b}q_{j,b'}}t^{b,b'}_{i,j} 
\langle
\hat{c}^{\dagger}_{i,b}\hat{c}^{}_{j,b'}\rangle_{\Psi_{0}}
\equiv\langle \hat{H}'_0  \rangle_{\Psi_{0}}
\end{eqnarray}
where we introduced the well known Gutzwiller loss factors \cite{Gutzwiller}
\begin{equation}
q_{i,b}=\frac{1}{n^0_{i,b}(1-n^0_{i,b})}\left( 
\sqrt{m_{i,\emptyset}m_{i,b}}+\sqrt{m_{i,\bar{b}}m_{i,d}}
 \right)^2
\end{equation}
and $\bar{b}$ is defined via  $\bar{1}\equiv 2$ and $\bar{2}\equiv 1$.

The lattice symmetry of the order parameter leads to further
simplifications. We introduce the majority and the minority orbital density
\begin{equation}
n^{0}_{\pm}\equiv n^{0}\pm\tau_0
\end{equation}
such that
\begin{equation}
\label{30}
n^{0}_{i,1}=n^{0}_{\pm}\,\,,\,\,n^{0}_{i,2}=n^{0}_{\mp}\,.
\end{equation}
The upper and lower signs in (\ref{30}), and in corresponding
 equations below, 
  belong to lattice sites  $i\in A$ and $i\in B$, respectively.
For the other local expectation values we find 
\begin{eqnarray}
m_{i,d}&=&m_{d}\,,\\
m_{i,1}&\equiv&m_{\pm}=n^{0}_{\pm}-m_{d}\,, \\
m_{i,2}&\equiv&m_{\mp}=n^{0}_{\mp}-m_{d}\,, \\
m_{i,\emptyset}&=&1-2n^{0}+m_{d}\,.
\end{eqnarray}
A similar notation is introduced for the $q$-factors
\begin{equation}
q_{i,1}\equiv q_{\pm}\,\,\,\, , \,\,\, q_{i,2}\equiv q_{\mp}
\end{equation}
where 
\begin{equation}
\label{33}
q_{\pm}\equiv\frac{1}{n^0_{\pm}(1-n^0_{\pm})}\left( 
\sqrt{m_{\emptyset}m_{\pm}}+\sqrt{m_{\mp}m_{d}}
 \right)^2\;.
\end{equation}
The expectation value (\ref{28}) splits into four components
\begin{equation}
\label{35}
E_0=q_{+}E^{++}+q_{-}E^{--}+\sqrt{q_{+}q_{-}}(E^{+-}+E^{-+})
\end{equation}
which belong to the four different hopping channels between majority (`$+$')
and minority (`$-$') states. In momentum space, the one-particle expectation 
values in (\ref{35}) can be written as
\begin{eqnarray}
\label{40}
E^{\omega \omega'}&=&\frac{1}{4}\sum_{{\bf k}}
\sum_{b,b'}\epsilon_{k}^{b b'}\left(M^{1}_{b,b'}
\langle\hat{c}^{\dagger}_{{\bf k},b}\hat{c}^{}_{{\bf
    k},b'}\rangle_{\Psi_0}+
\omega M^{2}_{b,b'}
\langle\hat{c}^{\dagger}_{{\bf k}+{\bf Q},b}\hat{c}^{}_{{\bf
    k},b'}\rangle_{\Psi_0}\right.\\ \nonumber
&&\left.+\omega' M^{3}_{b,b'}
\langle\hat{c}^{\dagger}_{{\bf k},b}\hat{c}^{}_{{\bf
    k}+{\bf Q},b'}\rangle_{\Psi_0}+\omega \omega' M^{4}_{b,b'}
\langle\hat{c}^{\dagger}_{{\bf k}+{\bf Q},b}\hat{c}^{}_{{\bf
    k}+{\bf Q},b'}\rangle_{\Psi_0}\right)
\end{eqnarray} 
where $\omega, \omega'$ represent the $+$ or $-$ signs and the 
coefficients $M^{\gamma}_{b,b'}$ are given as the elements of the matrices
\begin{eqnarray}
\tilde{M}^{1}&=&\left(\begin{array}{cc}
1&1\\
1&1
\end{array}
\right)\;\;\;\;\;,\;\;\;\tilde{M}^{2}=
\left(\begin{array}{cc}
1&1\\
-1&-1
\end{array}
\right)\;,\\ \nonumber
\tilde{M}^{3}&=&\left(\begin{array}{cc}
1&-1\\
1&-1
\end{array}
\right)\;\;\;,\;\;\tilde{M}^{4}=
\left(\begin{array}{cc}
1&-1\\
-1&1
\end{array}
\right)\;.
\end{eqnarray} 
 For the evaluation of the expectation values in (\ref{40}) we need 
to determine the one-particle wave function $|\Psi_0 \rangle$. Following
 reference \cite{springer,thul}, $|\Psi_0 \rangle$  is given as the ground 
state of the effective one-particle Hamiltonian 
\begin{equation}
\label{50}
\hat{H}_0^{\rm eff}=\hat{H}'_0-\eta\sum_i \exp{({\rm i}{\bf Q}{\bf
    R}_{i})}
(\hat{n}_{i,1}-\hat{n}_{i,2})\,,
\end{equation}
where $\hat{H}^{\prime}_0$ was introduced in (\ref{28}) and the term
proportional to the variational parameter $\eta$ allows to vary the order
parameter $\tau_0$.

In this work, we only aim to investigate the stability of the
orbitally unordered state. Therefore we just need to analyse the 
energy expression (\ref{35}) for small values of the order parameter
$\tau_0$. An expansion of the $q$-factors (\ref{33}) up to
second order in $\tau_0$,
\begin{equation}\label{50b}
q_{\pm}\approx \tilde{q}_0\pm \tilde{q}_1\tau_0
+\tilde{q}_2\tau_0^2 \,
\end{equation}
yields
\begin{equation}
\sqrt{q_{+}q_{-}}\approx \tilde{q}_0+\left(\tilde{q}_2-
\frac{\tilde{q}^2_{1}}{2\tilde{q}_0}  \right)\tau_0^2\,.
\end{equation}
Note, that the coefficients $\tilde{q}_{\gamma}$ are still functions
of $n^0$ and $m_d$. To leading order in $\tau_0$ the effective
Hamiltonian (\ref{50}) becomes
\begin{equation}
\label{55}
\hat{H}_0^{\rm eff}=\tilde{q}_0\sum_{{\bf k}}\sum_{b,b'}\epsilon^{b,b'}_{{\bf k}}
\hat{c}^{\dagger}_{{\bf k},b}\hat{c}^{}_{{\bf k},b'}-
\eta \sum_{{\bf k}}
\left(\hat{c}^{\dagger}_{{\bf k},1}\hat{c}^{}_{{\bf
    k}+{\bf Q},1}
-\hat{c}^{\dagger}_{{\bf k},2}\hat{c}^{}_{{\bf
    k}+{\bf Q},2}\right)
\end{equation}
since we can set $q_{i,b}=\tilde{q}_0$ in $\hat{H}'_0$,
eq. (\ref{28}). The one-particle Hamiltonian (\ref{55}) is
easily diagonalised numerically. This diagonalisation yields the
 coefficients in the quadratic expansion of (\ref{40})
\begin{eqnarray}
E^{\pm\pm}&=&E_0
+E_2\tau_0^2 \,,\\
E^{\pm\mp}&=&E'_{0}
+E'_{2}\tau_0^2\,,
\end{eqnarray}
and, consequently, of  the variational ground state energy   
\begin{equation}
E_{\rm var}=\tilde{q}_0 E_{0}^{\rm tot}+
\left(\tilde{q}_0 E_{2}^{\rm
  tot}
+\tilde{q}_2 E_{0}^{\rm tot}-E'_{0}\frac{\tilde{q}^2_{1}}{\tilde{q}_0} \right)
\tau_0^2
+U m_d \,.
\end{equation}
\begin{figure}[b]
\centerline{\includegraphics[clip,width=11cm]{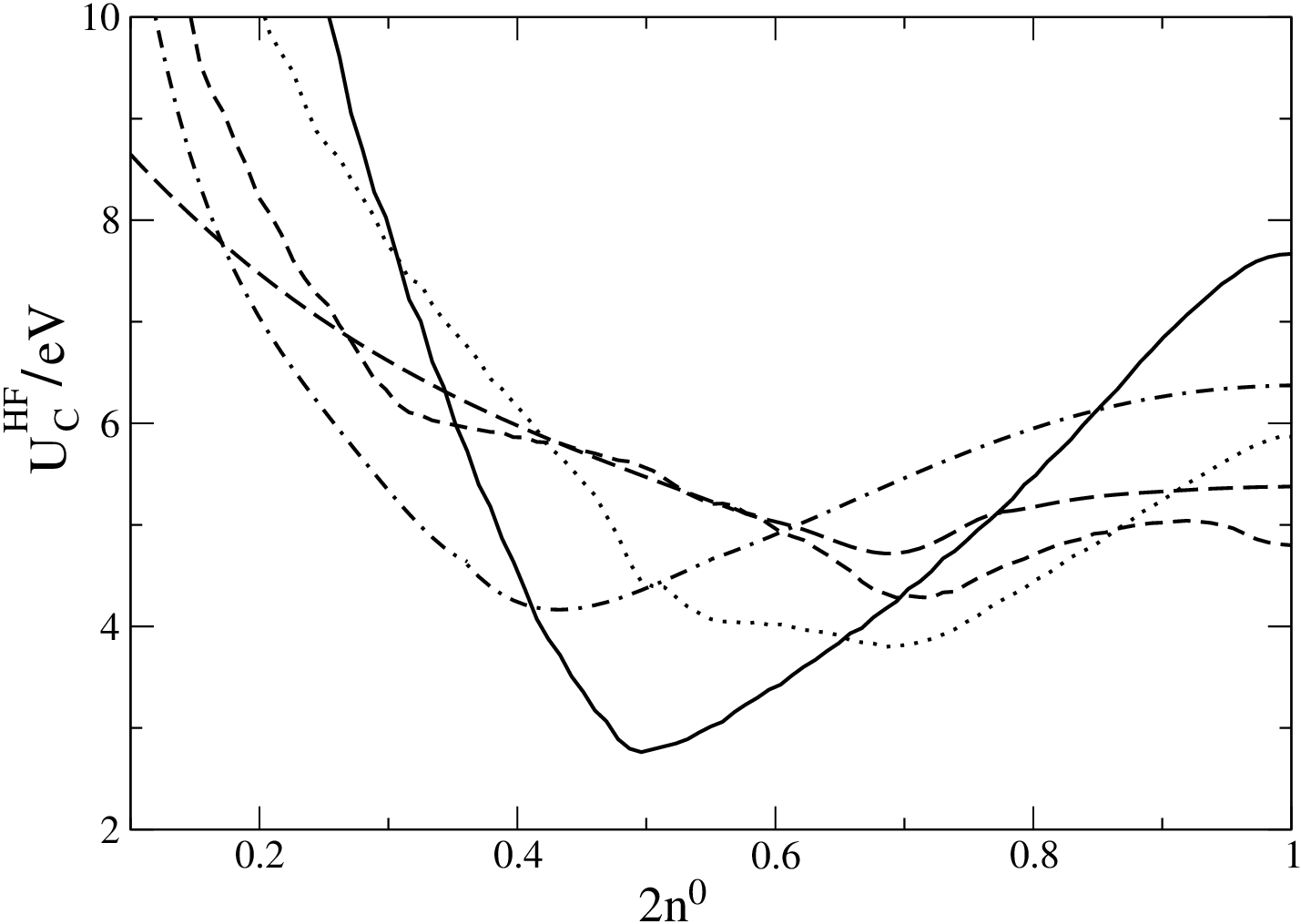}}
\caption{Critical interaction strength in Hartree Fock-theory for 
the orders $\tau^{y}_{R}$ (straight),  $\tau^{z}_M$ (dashed), 
$\tau^{z(x)}_{\Gamma}$ (long-dashed),
$\tau^{y}_M$ (dotted), and $\tau^{x}_X$ (dash-dotted). 
These phases are not stable within the Gutzwiller theory.}\label{fig2}
\end{figure}
Here, we introduced
\begin{eqnarray}
E_{\gamma}^{\rm tot}&\equiv&2(E_{\gamma}+E'_{\gamma})
\end{eqnarray}
for $\gamma=0$ or $2$. The minimisation of $E_{var}$ with respect to
$m_d$ can be carried out for $\tau_0=0$
 \begin{equation}
E_{0}^{\rm tot} \left. 
\frac{\partial  \tilde{q}_0}{\partial m_d}\right |_{m_d=\bar{m}_d}+U=0   
\end{equation}
and determines the optimum number  $\bar{m}_d$ of doubly occupied
sites as a function of $U$ and $n^0$.
This allows the expansion of  the
variational energy purely in terms of $\tau_0$ 
 \begin{equation}
\label{60}
E_{\rm var}=\tilde{q}_0 E_{0}^{\rm tot}+c_{\tau_0}\tau_0^2 \,.
\end{equation}
A negative sign of the coefficient
 \begin{equation}\label{60b}
c_{\tau_0}=\tilde{q}_0 E_{2}^{\rm
  tot}
+\tilde{q}_2 E_{0}^{\rm tot}-E'_{0}\frac{\tilde{q}^2_{1}}{\tilde{q}_0}
=c_{\tau_0}(U,n^0)
\end{equation}
in (\ref{60}) indicates the instability of the unordered state. Note
that a positive $c_{\tau_0}$ not necessarily proves the stability of the 
unordered state since it does not exclude first order
transitions.

In  Hartree-Fock theory a quadratic expansion of the 
ground-state energy leads to
\begin{equation}
\label{61}
E_{\rm var}^{\rm HF}=E_{0}^{\rm tot}+(E_{2}^{\rm tot}-U)\tau_0^2 \,,
\end{equation}
and the critical interaction strength  in this approach is therefore
 given as
\begin{equation}
\label{62}
U^{\rm HF}_{\rm C}=E_{2}^{\rm tot}\,.
\end{equation}
\begin{figure}[b]
\centerline{\includegraphics[clip,width=11cm]{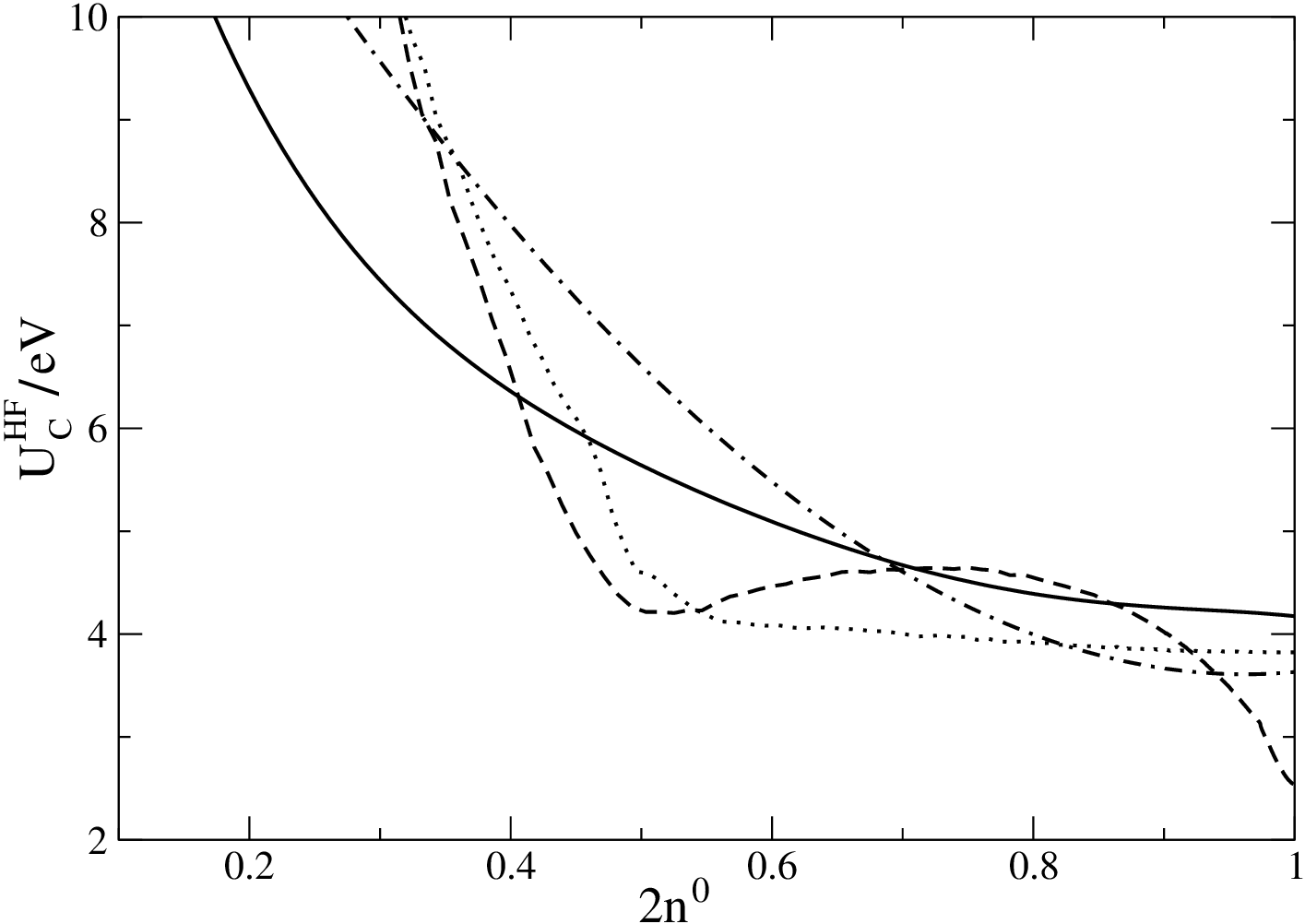}}
\caption{Critical interaction strength in Hartree Fock-theory for 
the orders $\tau^{y}_{\Gamma (X)}$ (straight),  $\tau^{z(x)}_R$ (dashed), 
$\tau^{x}_M$ (dotted), and $\tau^{z}_X$ (dash-dotted).
}
\label{fig3}
\end{figure}
\section{Results}
In figures~\ref{fig2} and~\ref{fig3} we show the critical interaction 
strength~(\ref{62}) in Hartree-Fock theory  as a function of density for the 
various types of orbital order introduced in section~\ref{sec:Hamilt}. 
Our data agree very well with those reported in reference~\cite{Takahashi}. 
Note that in Hartree-Fock  theory the critical interaction strength is 
finite for all densities $n^0> 0$ and diverges only in the limit 
$n^0\rightarrow 0$. 
 
The phases in figure \ref{fig2} are not stable 
within our correlated Gutzwiller approach for all densities and interaction 
parameters. This holds in particular for the order parameter 
$\tau_{R}^{y}$ which has surprisingly small critical values $U^{\rm HF}_{\rm C}$ 
around quarter filling $2n^0 \approx 0.5$.

\begin{figure}[t]
\centerline{\includegraphics[clip,width=11cm]{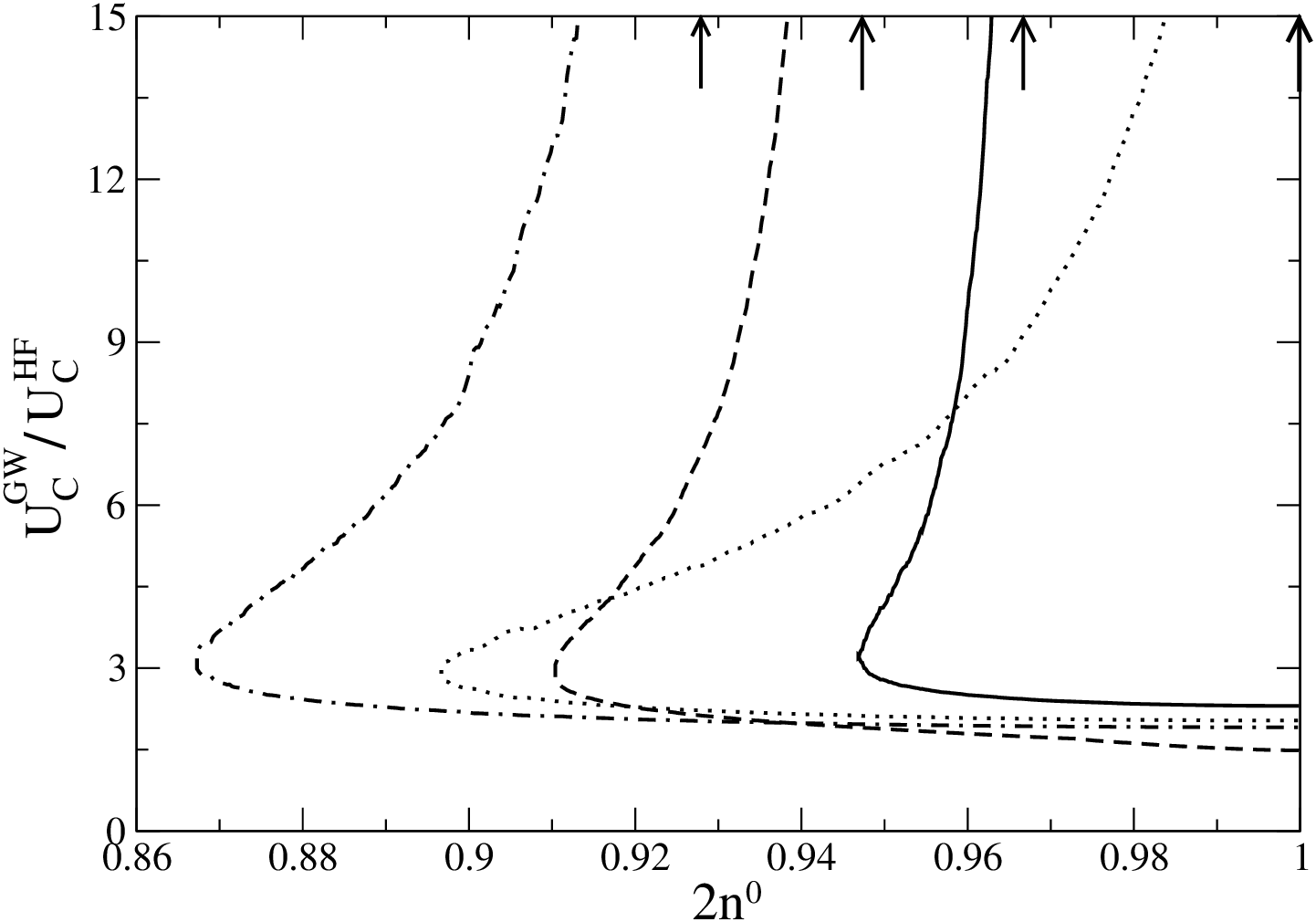}}
\caption{Critical interaction strength in Gutzwiller-theory for 
the orders $\tau^{y}_{\Gamma (X)}$ (straight),  $\tau^{z(x)}_R$ (dashed), 
$\tau^{x}_M$ (dotted), and $\tau^{z}_X$ (dash-dotted). The arrows 
 indicate the critical densities $2n_{\rm c}^0$ for $U\rightarrow 
 \infty$.}\label{fig4}
\end{figure}

For the four surviving phases (figure \ref{fig3}) we show the
ratio $U^{\rm GW}_{\rm C}/U^{\rm HF}_{\rm C}$ of the critical parameters 
 in Gutzwiller and Hartree-Fock theory as a function of density in 
 figure \ref{fig4}. 
Apparently, there is only a narrow window of densities 
around half filling $2n^0 \approx 1$ where orbital order occurs in the 
Gutzwiller theory. This is in stark contrast to the Hartree-Fock findings
 in figure~\ref{fig3}. 
For three of the phases the orbital order may disappear again if 
$U$ exceeds some second critical value 
$\tilde{U}^{\rm GW}_{\rm C}>U^{\rm GW}_{\rm C}$. This behaviour is also
different from the Hartree-Fock theory in which the orbital order 
is stable for all $U>U^{\rm HF}_{\rm C}$. Mathematically, the appearance
 of the second critical parameter $\tilde{U}^{\rm GW}_{\rm C}$ is due 
to the fact that $\tilde{q}_2$ in (\ref{50b}) has a minimum as a function 
of $m_{d}$. Hence, 
the coefficient $c_{\tau_0}$ may have one or two roots 
as a function of $U$, depending on the other parameters in (\ref{60b}). 
The appearance of the second transition is therefore a genuine many-particle 
effect.  Only one of the four orders, $\tau^{x}_M$, is unstable in the limit 
 $U\rightarrow \infty$ for all densities. For the three other orders we find 
critical densities $n_{\rm c}^0$ with a  stable order for all 
 $n^0>n_{\rm c}^0$ in the limit  $U\rightarrow \infty$. The values 
 of $n_{\rm c}^0$ are displayed by arrows in figure \ref{fig4}. 

The failure of Hartree-Fock theory to describe the orbitally ordered 
phases of the two-band model is not surprising. It is well known, 
 for example, that  Hartree-Fock theory also grossly overestimates 
  the stability of ferromagnetic ground states in the one-band Hubbard model 
\cite{fazekas}. Ferromagnetism in this model requires 
 very peculiar densities of states or very large local Coulomb interactions 
\cite{ulmke}. The simple Stoner criterium of the Hartree-Fock theory, however, 
predicts ferromagnetism for all densities and for small Coulomb interactions.

Orbital order in the two-band model at infinite $U$ has also been investigated 
 in reference \cite{oles}. In their work the authors report a complete 
disappearance of all types of orbital order seen in the Hartree-Fock theory.

\section{Summary}

We have reported results for orbital order in a two-orbital 
Hubbard-model without spin-degrees of freedom. In a previous work which 
was based on Hartree-Fock calculations, this system seemed to exhibit a 
surprisingly large variety of orbitally-ordered phases.
 In our study in this work we have employed the Gutzwiller 
variational theory which is known to be more reliable than the Hartree-Fock 
theory for systems with medium to strong local Coulomb interaction. 
Most of the phases found in the Hartree-Fock approach turned out to be  
unstable in the Gutzwiller-theory. 
Orbital order only appears for densities near half filling in our calculation. 
Unlike in Hartree-Fock theory, it may happen that an orbitally ordered phase 
which is stable for correlation parameters $U>U^{\rm GW}_{\rm C}$ becomes 
unstable again if $U$ exceeds a second critical value 
$\tilde{U}^{\rm GW}_{\rm C}>U^{\rm GW}_{\rm C}$. This second transition is a 
genuine many-particle effect.

Our findings show  that the stability of phases 
with broken symmetry for correlated electron systems can be
 grossly overestimated  by the Hartree-Fock mean-field theory. 
It is quite likely that similar problems appear in LDA+U calculations 
where the local Coulomb interaction is also treated on a Hartree-Fock level.


\end{document}